\documentclass[12pt]{article}
\usepackage{epsfig}

\def\beq{\begin{equation}}
\def\eeq#1{\label{#1}\end{equation}}
\def\eeqn{\end{equation}}

\def\beqa{\begin{eqnarray}}
\def\eeqa#1{\label{#1}\end{eqnarray}}
\def\eeqan{\end{eqnarray}}

\let\bar=\overbar

\def\etal{{\it et al.}}

\def\Dslash{\not{\hbox{\kern-4pt $D$}}}
\def\dslash{\not{\hbox{\kern-2pt $\del$}}}

\def\ee{e^+e^-}

\def\msb{{\bar{\ssstyle M \kern -1pt S}}}

\def\sax{\mbox{SAX J1808.4-3658}}
\def\be{\begin{equation}}
\def\ee{\end{equation}}
\def\lsim{\lower0.5ex\hbox{$\; \buildrel < \over \sim \;$}}

\def\dm{\mbox{$\dot{M}$}}
\def\msun{\mbox{$M_{\odot}$}}
\def\ra{\mbox{$R_{\rm A}$}}
\def\rco{\mbox{$R_{\rm c}$}}
\def\ro{\mbox{$R_0$}}
\def\rs{\mbox{$R_{\rm s}$}}

\def\dmmax{\mbox{$\dm_{\rm max}$}}
\def\dmmin{\mbox{$\dm_{\rm min}$}}
\def\fmax{\mbox{$F_{\rm max}$}}
\def\fmin{\mbox{$F_{\rm min}$}}
\def\etal{\mbox{\it et al.}}
\def\apj{Astrophys. J.\ }
\def\aap{Astron. Astrophys.\ }

\def\nat{\mbox{Nature\ }}
\def\sci{\mbox{Science\ }}
\def\Title#1{\begin{center} {\Large {\bf #1} } \end{center}}

\begin{document}

\Title{Strange star candidates}

\bigskip\bigskip


\begin{raggedright}  

{\it Ignazio Bombaci\index{Bombaci, I.}\\
Dipartimento di Fisica ``Enrico Fermi''\\
via Buonarroti, 2\\
I-56127 Pisa, ITALY}
\bigskip\bigskip
\end{raggedright}

\section{Introduction}
One of the most exciting aspects of modern astrophysics is the possible 
existence of a new family of compact stars, which are made entirely 
of deconfined {\it u,d,s} quark matter ({\it strange quark  matter} (SQM)). 
These strange quark matter stars are called in the scientific literature
{\it strange stars} (SS).   
They differ from neutron stars, where quarks are confined within neutrons, 
protons, and eventually within other hadrons (hadronic matter stars).   
The possible existence of SS is a direct consequence of the so called 
{\it strange matter hypothesis} \cite{witt}.   
According to this hypothesis, SQM (in equilibrium with respect to the weak 
interactions) could be the true ground state of matter.  
In other words, one assumes the energy per baryon of SQM (at the baryon 
density where the pressure is equal to zero) to be less than the lowest  
energy per baryon found in nuclei, which is about 930 {\rm MeV} for $^{56}$Fe. 
According to the strange matter hypothesis, the ordinary state 
of matter, in which quarks are confined within hadrons, is a metastable 
state \cite{witt,fj84}. The strange matter hypothesis does not conflict with 
the existence of atomic nuclei as conglomerates of nucleons, or with the 
stability of ordinary matter \cite{fj84,mads99,bomb2001}.  

From a basic point of view the equation of state for SQM should 
be calculated solving QCD at finite density. As we know, such a fundamental 
approach is presently not doable. Therefore one has to rely 
on phenomenological models. In this work, we use two simple phenomenological 
models for the equation of state (EOS) of strange quark matter.   
One is a model \cite{fj84} which is related to the MIT bag model for hadrons.  
The other is a model proposed by Dey {\it et al.} \cite{dey98}.  

\section{The mass--radius relation for compact stars} 
To distinguish whether a compact star is a neutron star or 
a strange star, one has to find a clear observational signature.
There is a striking qualitative difference in the mass--radius (MR)   
relation of SS with respect to that of neutron stars (see Fig.~1). 
For SS with ``small'' ($M << M_{max}$) gravitational mass, 
$M$ is proportional to $R^3$. 
In contrast, neutron stars have radii that decrease with increasing mass.  
This is a consequence of the underlying interaction between the stellar 
constituents which makes ``low'' mass SS self-bound objects
(see {\it e.g.} ref.\cite{book}) contrary to the case of neutron stars 
which are bound by gravity.    
As we know, there is a minimum mass for a neutron star 
($M_{min} \sim 0.1~M_\odot$). In the case of a strange star, there is 
essentially no minimum mass. 
As the central density $\rho_c \to \rho_s$ (surface density), a 
strange star (or better a strangelet for very low baryon number) is a 
self--bound system, until the baryon number becomes so low that finite size 
effects destabilize it.   


The transient X-ray burst source \sax\ was discovered in September 1996  
by the BeppoSAX satellite.   
Two bright type-I X-ray bursts were detected, each lasting less than 
30 seconds. Analysis of the bursts in \sax\ indicates that it is 4~kpc 
distant and has a peak X-ray luminosity of $6\times 10^{36}~$erg/s in its 
bright state, and a X-ray luminosity lower than $10^{35}~$erg/s in 
quiescence \cite{zand}. 
Coherent pulsations at a period of 2.49 milliseconds were discovered 
\cite{wijn}. 
The binary nature  of \sax\ was firmly established with the detection of a 
2 hour orbital period \cite{CM98} as well as with the optical identification 
of the companion star.     
\sax\ is the first pulsar to show both coherent pulsations in its persistent 
emission and X-ray bursts. 

A mass--radius (MR) relation for the compact star in \sax\ has been  
obtained by Li \etal\  \cite{Li99a} using the following two requirements.  
({\it i}) Detection of X-ray pulsations requires that the inner radius $\ro$ 
of the accretion flow should be larger than the stellar radius $R$. 
In other words, the stellar magnetic field must be strong enough to 
disrupt the disk flow above the stellar surface. 
({\it ii}) The radius $\ro$ must be less than the so-called co-rotation 
radius $\rco$, {\it i.e.} the stellar magnetic field must be weak enough 
that accretion is not centrifugally inhibited: 
$ \ro   \lsim  \rco = [GM P^2/(4\pi^2)]^{1/3}$.    
Here $G$ is the gravitation constant, $M$ is the mass of the star, 
and $P$ is the pulse period. The inner disk radius $\ro$ is generally 
evaluated in terms of the Alfv\'en radius $\ra$, at which the magnetic   
and material stresses balance \cite{BvdH91}: 
$\ro=\xi\ra=\xi[B^2R^6/\dm(2GM)^{1/2}]^{2/7}$, where $B$ and $\dm$ 
are respectively the surface magnetic field and the mass accretion 
rate of the pulsar, and $\xi$ is a parameter of order of unity 
almost independent \cite{li97} of $\dm$.   
Since X-ray pulsations in \sax\ were detected over a wide range of mass  
accretion rate (say, from $\dmmin$ to $\dmmax$), the two conditions 
({\it i}) and ({\it ii}) give $R\lsim \ro(\dmmax)< \ro(\dmmin)\lsim \rco$.  
Next, we assume that the mass accretion rate $\dm$ is proportional to the 
X-ray flux $F$ observed with RXTE. This is guaranteed by the fact that the 
X-ray spectrum of \sax\ was  remarkably stable and there was only slight 
increase in the pulse amplitude when the X-ray luminosity varied by a factor 
of $\sim 100$ during the 1998 April/May outburst \cite{gil98,cui98,PC99}.  
Therefore, Li \etal\  \cite{Li99a} get the following upper limit of the 
stellar radius:   $ R < (F_{min}/F_{max})^{2/7} \rco$, or 
\be  
      R < 27.5  \bigg({{F_{min}}\over{F_{max}}}\bigg)^{2/7}  
                 \bigg({{P}\over{2.49~{\rm ms}}}\bigg)^{2/3} 
                 \bigg({{M}\over{M_\odot}}\bigg)^{1/3} ~{\rm km},  
\label{eq:MR-sax}
\ee
where $\fmax$ and $\fmin$ denote the X-ray fluxes measured 
during X-ray high- and low-state, respectively, $\msun$ is the 
solar mass.  Note that in writing inequality (1) it is assumed 
that the pulsar's magnetic field is basically dipolar 
{\footnote{~see ref.\cite{Li99a} for arguments to support this hypothesis. 
See also ref.\cite{PC99} for a study of the influence on the MR relation for 
\sax\ of a quadrupole magnetic moment, and of a {\it non-standard} 
disk--magnetosphere interaction model.}}.  

\begin{figure}[htb]
\includegraphics[scale=0.5, angle=90]{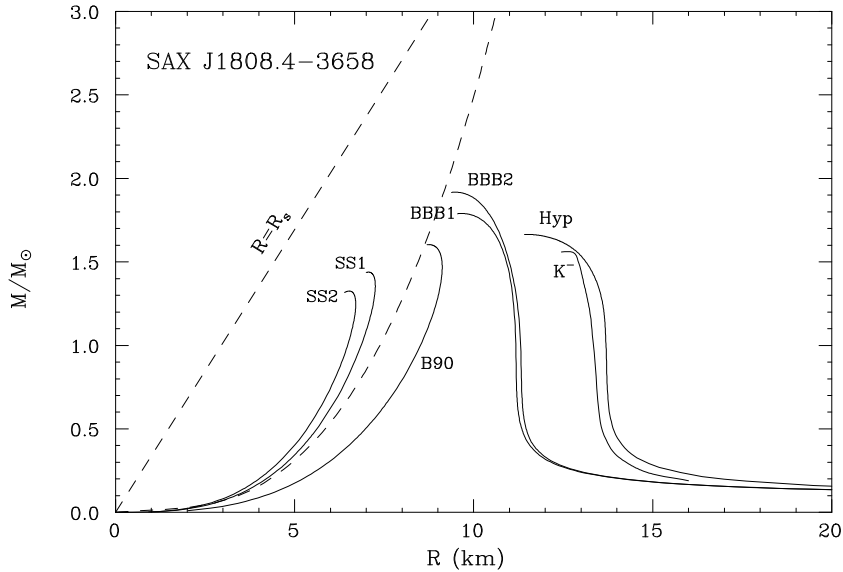}
\vspace{-0.8truecm}
\caption{Comparison of the MR relation of SAX J1808.4 -3658 
determined from RXTE observations with theoretical models of neutron 
stars and of SS.  The solid curves represents theoretical MR 
relations for neutron stars and strange stars.}
\label{fig:fig1}
\end{figure}

Given the range of X-ray flux at which coherent pulsations were 
detected, inequality (1) defines a limiting curve in the MR plane 
for SAX J1808.4-3658, as plotted in the dashed curve in Fig.~1. 
The authors of ref.\cite{Li99a} adopted the flux ratio    
$\fmax/\fmin\simeq 100$ from the measured X-ray fluxes with the RXTE   
during the 1998 April/May outburst \cite{cui98,PC99}.  
The dashed line $R = \rs \equiv 2GM/c^2$ represents the Schwartzschild 
radius  - the lower limit of the stellar radius to prevent the star collapsing 
into a black hole.  Thus the allowed range of the mass and radius of 
\sax\ is the region confined by these two dashed curves in Fig.~1. 

In the same figure, we report the theoretical MR relations (solid curves) 
for neutron stars given by some recent realistic models for the EOS of 
dense matter (see ref.\cite{Li99a} for references to the EOS models). 
Models BBB1 and BBB2 are relative to ``conventional'' neutron stars 
({\it i.e.} the core of the star is assumed to be composed by an 
uncharged mixture of neutrons, protons, electrons and muons in 
equilibrium with respect to the weak interaction).   
The curve labeled Hyp depicts the MR relation for a neutron 
star in which hyperons are considered in addition to nucleons as hadronic 
constituents. The MR curve labeled $K^-$ is relative to neutron stars  
with a Bose-Einstein condensate of negative kaons in their cores.   
It is clearly seen in Fig.~1 that none of the neutron star MR curves 
is consistent with \sax. Including rotational effects will shift 
the $MR$ curves to up-right in Fig.~1 \cite{dtb98}, and does not 
help improve the consistency between the theoretical neutron star models 
and observations of \sax.   
Therefore \sax\ is not well described by a neutron star model. 
The curve B90 in Fig.~1 gives the MR relation for SS described by the 
schematic EOS of ref. \cite{fj84} for massless non-interacting quarks 
with B = 90 MeV/fm$^3$.  
The two curves SS1 and SS2  give the MR relation for SS calculated with 
the EOS of Dey {\it et al.}\cite{dey98} for two parameterizations which 
give absolutely stable SQM according to the strange matter hypothesis. 
Clearly a strange star model is more compatible with \sax\ than 
a neutron star one.  

Stringent constraints on the MR relation have been also obtained for the 
compact star in the X-ray source 4U~1728-34 
($M < 1.0~M_\odot$ and  $R < 9 ~{\rm km}$) \cite{Li99b},   
for the isolated compact star RX~J1856-37  
($M = 0.9 \pm 0.2~M_\odot$ and  $R = 6 ^{+2}_{-1}~{\rm km}$) \cite{Pons01} 
(see also \cite{j1856})   
and for the X-ray pulsar Her X-1 
($M = 1.1$ -- $1.8~M_\odot$ and  $R = 6.0$ -- $7.7~{\rm km}$) \cite{dey98}. 
Clearly it is very difficult to model the MR relation for these compact  
objects (see {\it e.g.} Fig. 1) using any realistic EOS for neutron 
star matter. 

\section{Astrophysical implications and final remarks}
If the strange matter hypothesis is true, then a neutron star could 
``decay'' to a strange star (NS$\rightarrow$SS conversion)  
once a ``seed'' of SQM forms in the neutron star's core or it comes from 
the galactic space \cite{afo86}. 
The conversion of the whole star occurs in a very short time \cite{hb88}, 
in the range  1~ms -- 1~s, and liberates a total energy \cite{bd2000} 
of a few  $10^{53}$ erg.   
The NS$\rightarrow$SS conversion has been proposed \cite{bd2000} 
as a possible energy source to power $\gamma$-ray bursts at 
cosmological (redshift $z \sim $ 1 -- 3) distances \cite{kulk}.     

Strange stars are the natural site for a color superconducting state 
of quark matter \cite{nard01,krishna}.
Particularly, there could be a region inside a strange star where quark 
matter is in a crystalline (``LOFF''\cite{LOFF}) superconducting 
phase \cite{nard01,krishna}.  
This raises the possibility to successfully model pulsar glitches with 
strange stars \cite{krishna}. 

A very unpleasant consequence of the strange matter hypothesis could be 
the possible formation of stable negatively charged strangelets 
during heavy ion collisions at RHIC or at LHC. In fact, it has been 
pointed out \cite{disaster} that these ``dangerous'' negatively charged 
strangelets may  trigger the disruption of our planet. 
Luckily, there are various theoretical as well as experimental 
arguments \cite{disaster} to rule out this ``Disaster Scenario''.

The main result of the present work ({\it i.e.} the likely existence 
of strange stars) is based on the analysis of observational data  
for the X-ray sources SAX J1808.4-3658. 
The interpretation of these data is done using {\it standard} models for 
the accretion mechanism, which is responsible for the observed phenomena.     
The present uncertainties in our knowledge of the accretion mechanism, 
and the disk--magnetosphere interaction,  do not allow us to definitely 
rule out the possibility of a neutron star for this X-ray source.   
For example, making {\it a priori} the {\it conservative} assumption that 
the compact object in \sax\ is a neutron star, and using a MR relation similar 
to our eq.(1), Psaltis and Chakrabarty \cite{PC99} try to constrain 
disk--magnetosphere interaction models or to infer the presence of a 
quadrupole magnetic moment in the compact star. 

 \sax, 4U~1728-34, RX~J1856-37, and Her X-1 are not the only X-ray sources 
which could harbour a strange star. Recent studies have shown that the compact 
objects associated with the X-ray burster 4U 1820-30 \cite{Bomb97} and   
the bursting X-ray pulsar GRO J1744-28 \cite{Cheng} are likely strange star 
candidates.  
For each of these X-ray sources (strange star candidates) the conservative 
assumption of a neutron star as the central accretor would require 
some particular (possibly {\it ad hoc}) assumption about the nature of the 
plasma accretion flow  and/or the structure of the stellar magnetic field. 
On the other hand, the possibility of a strange star gives a simple 
and unifying  picture for all the systems mentioned above. 


\end{document}